# Enhanced Security for Cloud Storage using File Encryption

Debajyoti Mukhopadhyay, Gitesh Sonawane, Parth Sarthi Gupta, Sagar Bhavsar, Vibha Mittal

Department of Information Technology
Maharashtra Institute of Technology
Pune 411038, India
{debajyoti.mukhopadhyay, gitesh007, parthsarthigupta17, sagarbhavsar007, vibha.m807}@gmail.com

*Abstract-* **Cloud computing is a term coined to a network that offers incredible processing power, a wide array of storage space and unbelievable speed of computation. Social media channels, corporate structures and individual consumers are all switching to the magnificent world of cloud computing. The flip side to this coin is that with cloud storage emerges the security issues of confidentiality, data integrity and data availability. Since the "cloud" is a mere collection of tangible super computers spread across the world, authentication and authorization for data access is more than a necessity. Our work attempts to overcome these security threats. The proposed methodology suggests the encryption of the files to be uploaded on the cloud. The integrity and confidentiality of the data uploaded by the user is ensured doubly by not only encrypting it but also providing access to the data only on successful authentication.**

## I. INTRODUCTION

"Cloud" is a term used for a virtual collection of computing resources. A wide range of benefits are offered to consumers using cloud computing: availability of a huge array of software applications, seemingly unlimited storage, access to lightning-fast processing power and the ability to easily share information across the globe. A user can access all of these benefits through his browser any time once he has access to the Internet. In the early 1990s, a large ATM network started being referred to as "cloud" [1]. The term appeared once again about twelve years ago with the arrival of Amazon's web-based services. Cloud computing allows consumers and corporate structures to use all the applications offered by the cloud without the extra effort of installation and also offers access to their personal files from any computer with Internet access.

Cloud computing is a complex infrastructure of software, hardware, processing, and storage, all of which are available as a service. It is comprised fundamentally of applications running remotely (popularly referred to as "in the cloud") which is made available to all its users. This technology offers access to a large number of sophisticated supercomputers and their resultant processing power, connected at numerous locations around the world, thus offering speed in the tens of trillions of computations per second.

Cloud promises tangible cost savings and speed to customers. Using the technology of cloud, a company can rapidly deploy applications where expansion and contraction of the core technology components can be attained with the high and low of the business life cycle. This can be achieved with the help of cloud enablers, such as virtualization and grid computing, that allow applications to be dynamically deployed onto the most suitable infrastructure at run time. It's worth noting that while this might appear alluring, there remain issues of reliability, portability, privacy and security.

## II. CLOUD SERVICES

A lot of research has been done on the potential of cloud and the services that cloud computing can and could offer. In essence though, these services can be categorized into four main sections: Storage as a Service (StaaS), Platform as a Service (PaaS), Infrastructure as a Service (IaaS) and Software as a Service (SaaS).

*A. Storage as a Service*

Cloud offers a storage space that is enormous, seemingly endless, and growing every day. Storage as a Service (StaaS) enables cloud applications to scale beyond their limited servers. Cloud storage systems are expected to meet several laborious requirements for maintaining users' data and information, including high availability, reliability, performance, replication and data consistency. In addition to those services, consumers are also relieved of their responsibility to own and maintain their own computer storage as cloud vendors offer them the choice of storing their information in the cloud which is accessible whenever they want. Unfortunately, due to the conflicting nature of the requirements of cloud services, no one system implements all of them together.

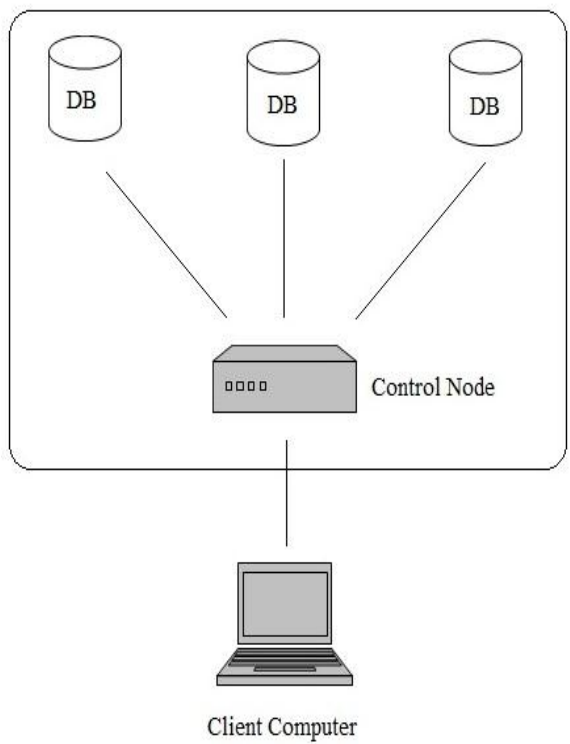

Fig. 1. A typical cloud storage system.

*B. Software as a Service*

For users, SaaS is the most popular choice. SaaS offers benefits that applications running locally on your own computer can never offer. Using cloud, a user potentially has access to any application that he can never own from any computer supporting a browser. The ever increasing number of SaaS applications encourages users to create and collaborate on projects with people spread around the globe simply by uploading their data on the cloud where anyone can have access to it.

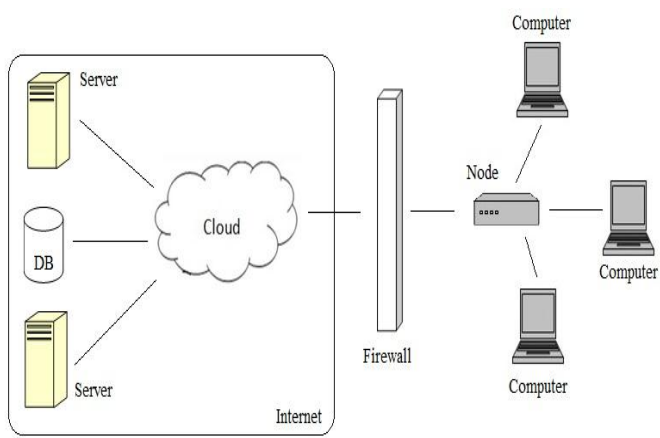

Fig. 2. Concept of software as a service

*C. Platform as a Service*

In this model, the user creates the software using tools and/or libraries provided by the cloud. The user controls the software deployment and configuration settings. Facilities offered by PaaS help the deployment of applications without the cost and complexity of buying and managing the underlying hardware and software. PaaS facilities also include enhancing the application life cycle processes like application design, application development, testing and deployment.

*D. Infrastructure as a Service*

IaaS is a model in which an organization outsources the equipment used to support operations such as storage, hardware equipment, servers and networking components. The service provider owns the equipment and is responsible for housing, running and maintaining it. The client typically pays on a per-use basis.

### III. SECURITY ISSUES

Companies are rapidly moving onto cloud because they can now use the best resources available on the market in the blink of an eye and also reduce their operations' cost drastically. But as more and more information is moved to the cloud the security concerns have started to develop.

- Data breaching is the biggest security issue. A capable hacker can easily get into a client side application and get into the client's confidential data. [2]
- Inefficient and flawed APIs and interfaces become easy targets. IT companies that provide cloud services allow third party companies to modify the APIs and introduce their own functionality which in turn allows these companies to understand the inner workings of the cloud.[2]
- Denial of Service (DoS) is also a major threat wherein the user is granted partial or no access to his/her data. Companies now use cloud 24/7 and DoS can cause huge increase in cost both for the user and service provider.
- Connection eavesdropping means that a hacker can scan your online activities and reproduce/replay a particular transmission to get into your private data. It can also lead to the user to illegal or unwanted sites.
- Data loss is also another issue. A malicious hacker can wipe out the data or any natural/man-made disaster can destroy your data. In such cases having an offline copy is a big advantage. Carelessness of the service provider can also lead to data loss.[3]
- Compatibility between different cloud services is also an issue. If a user decides to move from one

cloud to another the compatibility ensures that there is no loss of data.
- Cloud can also be used for wrong purposes i.e. cloud abuse. Due to the availability of latest technologies on the cloud it can be used for high end calculations which cannot be done on a standard computer.[2],[ 3]
- Insufficient understanding of cloud technologies can lead to unknown levels of risk. Companies move to cloud because it provides substantial reduction in cost but if transfer is done without proper background learning, the problems that arise can be even greater.
- Insider theft in the form of a current or former employee, a contractor, etc who is able to use the data for harmful purposes.
- Safe storage of encryption keys is also a problem. Even if you are using encryption for enhanced security, safe keeping of the key becomes an issue. Who should be the owner of the key? User seems to be the answer but how diligent and careful can he/she be will decide the security of the data.

*A. Additional aspects concerning cloud security*

Minqui Zhou et al. and Kresimir Popovic et al. classify the various threats concerning cloud security in the following areas:
- **Access:** The aim of cloud services to provide information to the user from any place, at any time. As a web service cloud enables the user to access his/her data from anywhere and this is applicable to all the services being provided by it. The client should know where data is being stored. In a situation where the client asks the cloud service provider to delete his/her data, the data should be deleted. The cloud service provider should not withhold any information.
- **Control:** In a cloud, controlling the system and its use is important. The amount of data that is visible to any member of the service provider should be controlled. The visibility of the data defines the level of control.
- **Compliance**: Proper authorities need to define laws to govern the safekeeping of data on the cloud because clouds can cross multiple jurisdictions around the world. If a piece of data is stored in a different country and it contains sensitive data that is wanted by the authorities then do the rules apply on data that data?
- **Data Integrity**: Data integrity in simple terms means that the data is preserved and no changes are made without the user's permission. In cloud, data integrity is a fundamental requirement.
- **Audit:** This means to keep a simple check on the activities happening on the cloud. The presence of an auditing mechanism can maintain a log, list of events, etc. to help prevent breaches.
- **Privacy Breaches:** The cloud service provider should inform its users about any breach in security. The user has to the right to know what is happening in his/her space. How does the service provider take care of this?
- **Confidentiality**: This ensures that the user's data is kept secret. Confidentiality is one aspect of cloud storage security that will raise questions in a common users mind. Cloud as such is a public network and is susceptible to more threats, thus, confidentiality is very important.

IV. PROPOSED WORK

`In this section, we propose a framework which involves securing of files through file encryption. The file present on the device will be encrypted using password based AES algorithm. The user can also download any of the uploaded encrypted files and read it on the system.

The advantages of AES are many. AES is not susceptible to any attack but Brute Force attack. However, Brute Force attack is not an easy job even for a super computer. This is because the encryption key size used by AES algorithm is of the order 128, 192 or 256 bits which results in billions of permutations and combinations. AES is also much faster than the traditional algorithms like RSA. Thus, it makes a fine choice for protection of data on the cloud.

It is to be noted that the proposed system works only when a stable internet connection is available.

*A. File Upload*

The steps for the file upload process are explained in this section:

1. Accept user name and password from the user
    - If user is authenticated, establish connection with the cloud
    - Else, show authentication error
2. Ask user to select file to be uploaded onto the cloud
3. Ask the user to enter a password for the encryption process
4. Save this password and generate a key from this password
5. Apply the encryption algorithm
6. Upload the file on to the cloud
7. Ask the user if he wishes to delete the file once it is uploaded
    - Delete the file if the user selects the option for deletion
8. Disconnect connection with the cloud

Step 1 is authenticating the user. The methodology suggested by us aims to prevent every possible attack on the user data. The first step towards the same is authentication. The system will accept the user name and password from the user. Once the two are verified, the user is given access to his files.

Once user name and password are entered by the user, check for its validity. Only if entered name and password are valid, the system will establish a connection with the cloud. If the entered name and password are not valid, the system shows an error and rejects the user.

The next steps work on the condition that the user is authenticated and a working connection with the cloud has been established.

Selection of the file to be uploaded is done in step 2. The user can select any text file available in the memory of the machine he is currently using.

Once the file to be uploaded is selected, proceed to step 3. This step asks the user for a password for the encryption process. The user is recommended to use long passphrases as their passwords. This password is used for generation of a key. The details of this generation are described in the next step.

Step 4 is a very crucial step for the system. In this step, a key for the encryption process is generated. AES is a symmetric key algorithm, i.e., it uses the same key which is used for encryption to decrypt the data too. This key is generated from the password using a key generator function. We recommend the use of PBKDF2 (Password Based Key Generation Function 2). PBKDF2 uses iterations in the order of thousands. This added computation makes password cracking even more difficult. This process is called key stretching. It is to be noted that though keys used for AES algorithm are not susceptible to any known attack, there's a possibility for the password to be attacked by the Brute-force attack. Therefore, the user is highly recommended to use long passphrases for the generation of the key.

Thus, in this step the system saves the password once it is entered and generates a random key for encryption.

Step 5 is the encryption step. In this step, our encryption algorithm, that is, the AES algorithm is applied to the plain text to generate the cipher text. As mentioned before, AES is not susceptible to any known attacks. Thus the user can be rest assured that his data is safe from the various threats concerning cloud security. A user's data is doubly secured because one, a person can access the data only the entered user name and password are valid and two, even if the login password of the user is attacked, the uploaded is file is encrypted which can be decrypted only if the user enters the password which he entered during the encryption process. This ensures confidentiality. Also, since the uploaded data is encrypted, no modification can be made to the cipher text. This ensures data integrity.

Once the cipher text is generated, upload the encrypted file to the cloud. This is the sixth step of your process.

The seventh step is concerned with the deletion of the original plain text file from the memory of the machine. The user is given an option to delete the original file once it is uploaded to the cloud. If the user does not wish to do so, he can select the second option of retaining the original file. We recommend that the original file should be deleted. This ensures that no unauthorized access will be made to the plain text file stored in the machine.

If the user selects the delete option, the system will delete the original file from the machine.

Once the cipher text file is successfully uploaded on the cloud and the user has no more files to be uploaded, the system logs out of the user account and disconnects the established connection with the cloud.

The process of file upload can be shown diagrammatically as:

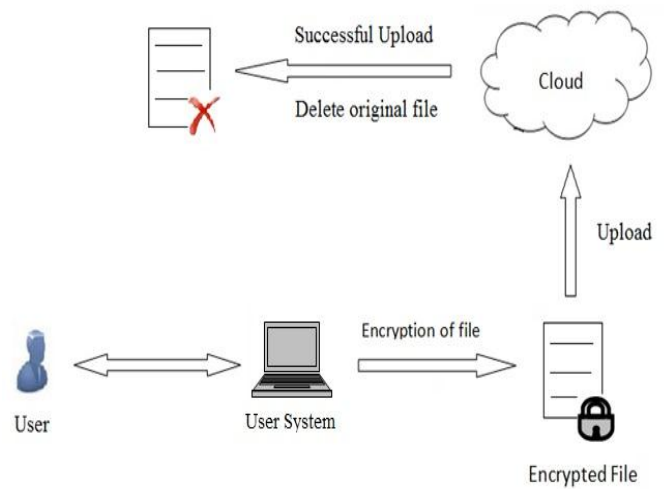

Fig. 3. File upload

*B. File Download*

The steps for file download process are explained in this section:

Step 1 is same as step 1 of file upload. The identity of the user is authenticated in this step.

In the second step, the array of files that the user has uploaded on the cloud so far is displayed. The user is asked to select one of the files from the list.

In step 3, the user is asked to enter the password which he had entered during the encryption of the file

Validation of this entered password is done in step 4. The cipher text file uploaded by the user will only be decrypted and downloaded if the password entered is same as the password entered during file encryption. This is the reason that the password is saved during the encryption process, that is, the stored password is used to validate the entered password. As, mentioned before, AES algorithm is a symmetric key algorithm. Therefore, it needs the same key to encrypt and decrypt the data. This is possible only if the same password is entered into the key generator function to generate the key.

Once, the entered password is validated, use the password to generate the decryption key using the key generator function. If the password is not validated, show an error message and reject the password.

In step 5, using the generated key, use the AES algorithm to decrypt the uploaded cipher text.

Save this decrypted plain text in the user machine's memory. This is done in step 6.

In step 7, the user is asked if he wishes to delete the cipher text file uploaded on the cloud. If the user chooses to do so, delete the encrypted file from the file

Once the user does not wish to download any more files from the cloud, log out of the user account and disconnect the established connection with the cloud. This is the last step of the download process.

1. Accept user name and password from the user
   - If user is authenticated, establish connection with the cloud
   - Else, show authentication error
2. Ask user to select file to be downloaded
3. Ask the user to enter a password for the decryption process
4. Check the validity of this password
   - If the password entered is valid, generate a key
   - Else show an error message and reject password
5. Apply the decryption algorithm
6. Download the file from the cloud
7. Ask user if he wants to delete the uploaded encrypted file
   - Delete the encrypted file from the cloud if user selects the delete option
8. Disconnect connection with the cloud

The process of file download can be shown diagrammatically as:

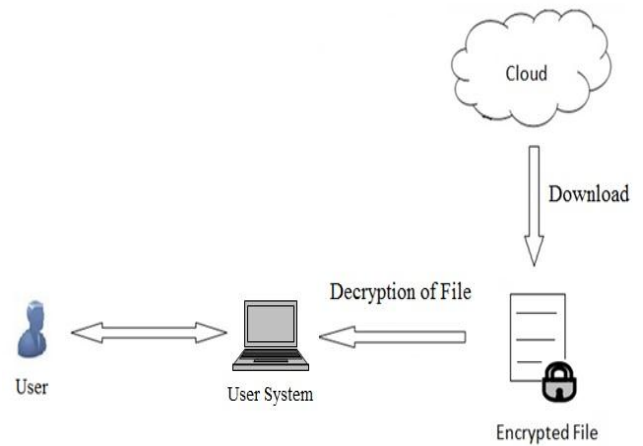

Fig. 4. File download

## V. FUTURE WORK

The proposed framework can be developed in the form of a mobile application using the various operating systems such as android, iOS, Windows and Symbian. It can also be integrated with any of the social networking sites to exchange data securely in its encrypted form. The algorithm can also be enhanced to not only encrypt text files but also audio and video files.

## VI. CONCLUSION

Cloud computing is an enigma anyone can get lost in. But just like any other technology, cloud computing is also a double edged sword. On one end lies the promise of lightning fast technology, a huge array of applications to use, seemingly unlimited storage space. On the other end lie various security threats which emerge with shared spaces such as breach of confidentiality, hampering of data integrity and non-availability of data. In this paper, we have proposed a framework which encrypts a file before it is uploaded on to the cloud. AES (Advanced Encryption Standard) is one of the most secure encryption algorithms and not many attacks are successful on data which is encrypted using AES. This proposal solves the problem of most, if not all, of the threats that data stored in the cloud faces. Our framework also suggests the use of login id and password to ensure authentic and authorized access to a user's data.

Thus, if used securely, cloud computing provides a user with amazing benefits and overcomes its only disadvantage of security threat.